\documentclass[rnote]{aa}
\usepackage{graphicx}
\usepackage{txfonts}

\def\lsim{\lower.5ex\hbox{$\; \buildrel < \over \sim \;$}}
\def\gsim{\lower.5ex\hbox{$\; \buildrel > \over \sim \;$}}

\begin{document}
\title{Properties of the propagating shock wave in the accretion flow around GX 339-4 in the 2010 outburst} 
\author{Dipak Debnath\inst{1}, Sandip K. Chakrabarti\inst{1,2} and Anuj Nandi\inst{1,3}}
\offprints{Dipak Debnath}
\institute{Indian Centre for Space Physics, Chalantika 43, Garia Station Rd., Kolkata, 700084, India
\and
S. N. Bose National Centre for Basic Sciences, Salt Lake, Kolkata, 700098, India
\and
On deputation from Indian Space Research Organization-HQ, Bangalore, India \\
   \email{dipak@csp.res.in; chakraba@bose.res.in; anuj@csp.res.in}}
   \date{Received ; accepted }
\abstract
{The black hole candidate GX 339-4 exhibited an X-ray outburst in January 2010, which is still 
continuing.
We here discuss the timing and the spectral properties of the outburst using RXTE data.}
{Our goal is to study the timing and spectral properties of GX 339-4 using 
its recent outburst data and extract information about the nature of the accretion flow.}
{We use RXTE archival data of the recent GX 339-4 outburst and analyze them with the 
NASA HEAsoft package, version 6.8. We then compare the observed quasi-periodic oscillation (QPO) 
frequencies with those from existing shock oscillation model and obtain the nature of evolution 
of the shock locations during the outburst.}
{We found that the QPO frequencies are monotonically increasing
from $0.102$ Hz to $5.69$ Hz within a period of $\sim 26$ days. We explain this evolution
with the propagating oscillatory shock (POS) solution and find the 
variation of the initial and final shock locations and strengths. 
The model fits also give the velocity of the propagating shock wave, which is responsible 
for the generation of QPOs and their evolutions, at $\sim 10$~m~s$^{-1}$. We
observe from the spectra that up to 2010 April 10, the object was in a hard state.
After that, it went to the hard-intermediate state. On April 18, it had a state transition
and went to the soft-intermediate state. On May 15, another state transition was observed
and the source moved to the soft state.}
{As in the previously fitted outburst sources, this source also showed the 
tendency of a rapidly increasing QPO frequency ($\nu_{QPO}$) in a viscous time scale, which 
can be modeled quite accurately. 
In this case, the shock seems to have disappeared at about $\sim 172$ Schwarzschild radii, 
unlike in the 2005 outburst of GRO J1655-40, where the shock disappeared behind the horizon.}

\titlerunning{QPO Evolution in GX 339-4 and the Accretion Disk Dynamics}
\authorrunning{Debnath, Chakrabarti, and Nandi}
\maketitle

\section{Introduction}

The source GX 339-4 is a well known stellar-mass Galactic black hole candidate. This bright variable
X-ray source was first observed during the survey period from October 1971 to January 1973 
by the MIT X-ray detector on-board the OSO-7 satellite in the energy range of 1-60 keV. GX 339-4, 
a transient low-mass X-ray binary (LMXB) system located at $(l,b) = (338^\circ.93,-4^\circ.27)$ 
(Markert et al. 1973) with R.A.=$17^h~02^m~49^s.36$ and Dec.= $-48^\circ~47'~22''.8$ (J2000). 
The optical spectroscopic study indicates that the mass function of the source is $M$ = $5.8\pm0.5~M_\odot$
and the distance $D$ = $6$~kpc (Hynes et al. 2003, 2004).

Since its discovery, GX 339-4 has undergone several outburst phases, during which the
source was observed in different wavebands to reveal the nature in multiple wavelengths (Liu et al.
2001, Homan et al. 2005). During the RXTE era (1996 onward), this source exhibited frequent 
X-ray outbursts (1998, 2002/2003, 2004/2005, 2006/2007) at a 2-3 years of interval with very
low luminosity states in between each episode. The complex outburst profile in each epoch
generally begins and ends in the low/hard state, which is quite common in other outburst 
candidate of black holes (e.g., GRO J1655-40, XTE J1550-564).
This general behavior is understood to be caused by sudden variation of viscosity in 
the system (Mandal \& Chakrabarti, 2010), which in turn causes the accretion rate of 
the standard Shakura-Sunyaev (1973) disk (hereafter referred to as the Keplerian rate)
to rise and possibly makes the inner edge move in. These transient black hole candidates show 
low and intermediate frequency quasi-periodic oscillations (QPOs) 
in their power density spectra. In general, during the 
rising hard state of the outburst the frequency of the QPO increases, whereas during the 
declining phase, the QPO frequency is gradually decreased.
The QPO evolution in these objects can be 
well understood through the propagating oscillatory shocks (POS; Chakrabarti et al. 2008, 2009). 
Though several studies of the evolution of the temporal and spectral states of GX 339-4 during 
the previous outbursts were carried out (Nowak et al. 1999, Belloni et al. 2005; 
Motta, Belloni \& Homan 2009), the underlying physical processes remained unclear.
Our attempt here is to see if the POS solution of our group can also explain the present outburst.
Note that for the traditional soft-X-ray transients 
with fast rise and exponential decay (FRED) lightcurve with typically long recurrence times,
there are so-called disk instability models (Cannizzo, 1993; Lasota, 1996) where matter also 
moves in owing to viscous processes. However these models do not address variations of QPOs.

Recently, after remaining in the quiescent state for three long years 
(except for a short spell of very weak activity in 2009 as observed in SWIFT/BAT), 
GX 339-4 became X-ray-active again on 2010 January 03, with a first detection by MAXI/GSC 
onboard HETE (Yamaoka et al. 2010). Immediately after the announcement of the X-ray trigger, 
RXTE started monitoring the source from 2010 January 12 (Tomsick, 2010). During the initial 
outburst phase, the source was in the low-hard state without any signature of QPO in the 
power density spectrum (PDS). In this outburst phase we first observed the QPO at $102$~mHz 
on 2010 March 22 (MJD 55277). After that, the QPO frequency monotonically increased to $5.69$~Hz 
until 2010 April 17 (MJD 55303). Afterward, QPOs were sporadically on and off (e.g., $5.739$~Hz, 
$5.677$~Hz and $5.913$Hz on April 18, 22 and 29 respectively), always remaining at about 
the same value. These sporadically appearing QPOs in PDS were observed until 2010 May 14. 
The observed QPOs in hard and hard-intermediate states (Homan \& Belloni, 2005) are of the 
`C' type and in soft-intermediate state are of the `B' type (van der Klis 2004, Casella et al. 2005).

The gradual increase of the QPO frequency in the rising phases of transient black hole 
and neutron star candidates have been known for a long time (e.g., Belloni \& Hasinger 1990, 
Belloni et al. 2002, Maitra \& Bailyn 2004). In the present context of studying outbursting 
black hole candidates, evolutions of QPOs during the 2005 outburst of GRO J1655-40 
(Chakrabarti, Debnath, Nandi \& Pal 2008, hereafter CDNP08) and the 1998 outburst of 
XTE J1550-564 (Chakrabarti, Dutta \& Pal 2009, hereafter CDP09) showed the monotonically 
increasing behavior in the rising phase like the one we find here in GX 339-4. 
However, while in GRO J1655-40 the QPO disappeared completely and then re-appeared after
about six months (Debnath et al. 2008) and in XTE J1550-564, $\nu_{QPO}$
started declining immediately after reaching maximum (CDP09), in the 
present case, $\nu_{QPO}$ began to stall at about $5.7-5.9$~Hz.
In CDNP08 and CDP09, it was shown that the oscillating shock that produces the 
QPOs was required to move in at a roughly constant speed of about $20~m~s^{-1}$. In both  
cases, a clear picture emerged about the system: it was found that while 
the low-angular momentum transonic flow (hereafter referred to as the sub-Keplerian matter, see, 
e.g. Chakrabarti, 1990) was always present even in the quiescent state, the Keplerian disk moved 
in closer to the black hole in the rising phase and then receded far away in the declining phase. 
This picture was corroborated when even the hardness-intensity diagram was reproduced 
(Mandal \& Chakrabarti, 2010) with this consideration.

We here examine the nature of the rising phase of GX 339-4
and show how it changed from hard state to soft state via a short-lived hard-intermediate
and soft-intermediate state. We also show that the evolution of QPOs can be understood by the 
POS solution as in the case of other outbursts. In the next section, 
we present the observational results of GX 339-4 since it
showed evidences of the outburst in January, 2010. 
We also fit the QPOs using POS solution presented in CDNP08 and CDP09.
Finally, in Sect. 3, we present the concluding remarks.

\section {Observational results and analysis}

We now present the timing and spectral properties of the GX 339-4 X-ray outburst using the 
archival data of the RXTE PCA and ASM instruments. We use the standard RXTE data analysis software 
package HEAsoft 6.8. For RXTE/ASM (Levine et al. 1998), one day averaged 
archival data of the different energy bands (2-3, 3-5, 5-12 \& 2-12 keV) were downloaded 
and analyzed. For the PCA data (Jahoda et al., 1996) analysis, we mainly use the most 
stable and well conditioned proportional counter unit 2 (PCU2) data (all six layers).
Background spectra were made using FTOOLS {\it runpcabackest} task 
and the most recent bright source model. The task {\it pcarsp} was used
to generate the PCA response file. 
For the timing analysis, we used the PCA Event mode data with a maximum timing resolution
of $125\mu s$, and for the spectral analysis we used the PCA `standard 2' data. 
In the entire PCA data analysis, we did not include the deadtime corrections 
because the counts are not very high, the maximum rate being around $1200$ cts/s. We verified
that the error caused by this is at most $4\%$.
In Fig. 1(a-b), we present the ASM lightcurve and the
hardness ratio as a function of days. Sudden changes in slope on MJD 55296, MJD 55304, 
and MJD 55331 are indications of the state transitions.
From the hardness ratio (Fig. 1b) we see that the spectrum was hard till MJD 55296. At 
MJD 55304 the spectrum became softer very quickly. This is the so-called hard-intermediate 
state (Homan \& Belloni 2005). After that, the ratio remained almost constant. Sporadic QPOs 
(see below) in this state indicate that the object is in the soft-intermediate state. Around 
2010 May 15 (MJD 55331), the source moved to the soft state, with a sharp fall in the count rate in the 
4-15 keV energy range. In Fig. 2 we show the total counts ($2-20$ keV) as a function of the 
hardness ratio $HR=(6-20~keV)/(2-6~keV)$ from 2010 January 12 (MJD 55208) until 
August 14 (MJD 55422). 
Four phases are clear: in the range A to B, the object is in the
hard state, in the range B to C, the object is in the hard-intermediate state,
C to D, the object is in the soft-intermediate state and beyond D, the object is in
the soft state. A detailed physical picture will be discussed below in Sect. 2.2.

\subsection{Timing analysis}

We analyze $114$ observational IDs from 2010 January 12 (MJD = 55208) to 2010 July 8 
(MJD = 55385). For the timing analysis, we use RXTE ASM and PCA public data. In the study of
the temporal properties of any black hole candidate, finding quasi-periodic oscillations (QPOs)
in power density spectra (PDS) is as important as the observation of the photon count/flux 
variations. To generate the PDS, we used the ``powspec" task of XRONOS package with a 
normalization factor of `-2' to have the `white' noise subtracted rms fractional variability 
on 2-15 keV (0-35 channels) PCU2 lightcurves of $0.01$ sec time bins. The power obtained 
has the unit of rms$^2$/Hz. Quasi-periodic oscillations are generally of a Lorentzian type 
(Nowak 2000, van der Klis 2005) and thus these are fitted with model Lorentzians. In Fig. 3a 
we show the variation of the QPO frequency in this period. The monotonically increasing 
frequency (starting from $102$~mHz observed on March 22 to $5.69$~Hz observed on April 17) 
as in  GRO J1655-40 and XTE J1550-564 (CDNP08, CDP09) motivated us to fit it with the same 
POS solution as used in CDNP08 and CDP09. In this solution, at the onset of the outburst, 
a shock wave moves toward the black hole, which oscillates either because of resonance 
(cooling time $\sim$ infall time; Molteni, Sponholz \& Chakrabarti, 1996) or because of the fact
that the Rankine-Hugoniot relation is not satisfied (Ryu, Chakrabarti \& Molteni, 1997)
to form a steady shock.
The QPO frequency is obtained from the inverse of the in-fall time scale from the post-shock
region. The oscillation of the shock produces the oscillation of hard X-ray intensity, because 
the post-shock flow behaves like a Compton cloud that intercepts a variable number of soft 
photons during oscillations. The governing equations are in CDNP08 and will not be repeated 
here. By fitting the QPO frequencies with POS solution  we find that the compression ratio 
$R=\rho_-/\rho_+$ monotonically goes down from $R_0=4$ (strongest shock) to $\sim 1$ (weakest 
shock) through the relation $1/R\rightarrow 1/R_0 + \alpha  (t_d)^{2.15}$, where, $t_d$ is the 
time in days (taking the first day of the QPO observation as the $0^{th}$ day, i.e., from 
March 22). Here, $\alpha$ is a constant that determines how rapidly the shock strength decreases 
with time. The value of $\alpha$ is obtained by the constraint that on the last day (i.e., after 
$t_r\sim 26$ on April 17) the  QPO  was observed, the shock became the weakest ($R\sim 1$). Using 
this condition that $R\rightarrow 1$ on $t_d=t_r$, we obtain $\alpha$ at $6.8 \times 10^{-4}$ 
in the present case. In Fig. 3b we show the fitted shock strength and the shock location as 
a function of day, where $0$ is March 22 when the QPO was first detected.
According to our fit, the shock started at $r \sim 1500$ Schwarzschild radii ($r_g=2GM/c^2$)
and disappeared on the $26^{th}$ day (April 17), when it was at $172$ Schwarzschild radii. 
The shock wave is found to move toward the black hole at a constant velocity of $\sim 10~m~s^{-1}$,
somewhat slower that the other two members, namely, GRO J1655-40 and XTE J1550-564 where the velocity
was about twice as high. After that, the QPO is seen sporadically, but the frequency remains about 
the same. This is because when the shock is weakest, moving inward cannot reduce its strength
any more. During this period, the Keplerian matter also moves in, increasing 
the softness of the spectrum (Fig. 1). In the soft-intermediate state, the Keplerian rate rises
to become comparable to the sub-Keplerian rate, while in the soft state (i.e., current
spectral state) the Keplerian rate dominates. The net duration of the soft-intermediate
and soft states would therefore depend on how long the viscosity remains sufficient high  
to maintain a Keplerian flow. Thus the prediction of the net duration of the outburst is not 
easy. Detailed modeling is required to predict the possible duration of any outburst.

\subsection{Spectral analysis}

For the spectral study we use $3-25$ keV PCA `standard-2' data of PCU2 and XSPEC 
(version 12.5) package. For all observations we kept the hydrogen column density 
($N_H$) fixed at $5 \times 10^{21}$ (Mota et al., 2009), using the absorption model 
{\it wabs}. All the spectra were fitted with standard disk blackbody and power-law models.
For each spectrum one Gaussian line at $\sim 6.5$~keV was used. For the best fit, we also 
added $1.0\%$ systematic error to the full spectrum. In Fig. 4(a-d) we show the evolution 
of photon counts and the photon index. We show the results from March 5 onward. 
Panel (a) shows very hard photons in the $15-30$ keV range, which could come only from 
Comptonization, (b) shows the photon counts of the intermediate energy range $4-15$ keV, 
which is monotonically increasing, and the panel (c) shows very soft photons ($2-4$ keV), 
which could come only from the blackbody component, i.e., the Keplerian disk.
Panel (d) shows the photon index $\Gamma$. Note that QPOs are observed immediately 
after a `kink' occurring on MJD 55274, the physics which is still unclear. 
$\Gamma$ remains less than $2$, and the very hard photons increase until MJD 55296. 
We may assume that the object is in the hard state. After MJD 55296, the hard photon 
count rapidly diminishes and the soft photon count rapidly increases. In this phase, 
the Keplerian rate increases and becomes comparable to the sub-Keplerian rate. 
As Chakrabarti \& Titarchuk (1995) pointed out, the spectral index becomes very 
sensitive to the Keplerian rate when the rates are comparable at around $\sim 0.5-1$ 
Eddington rate. This is the intermediate state. The index rapidly increased to about 
$2.5$ on MJD 55304.7 and then remained almost constant, perhaps because the Keplerian 
disk rate had reached its peak and was sustained by the viscosity. In future, when 
the viscosity is reduced, the Keplerian rate would decrease and the object will gradually 
go back to the hard state. With a very low Keplerian rate, the inner part of the disk 
evaporates, which is equivalent to saying that its inner edge has receded.

\begin{figure}
\vbox{
\vskip 0.6cm
\centerline{
\includegraphics[scale=0.6,angle=0,width=8truecm]{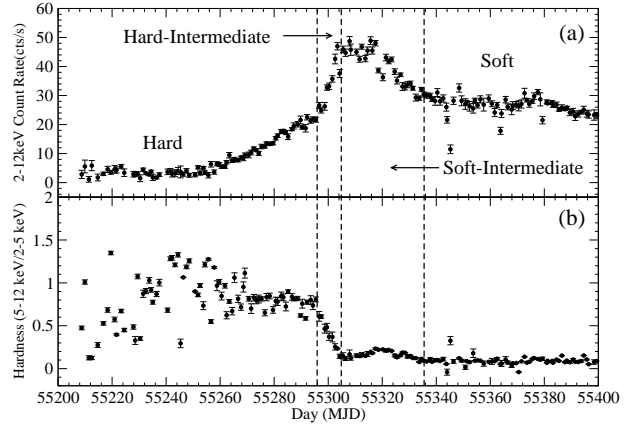}}
\vspace{0.0cm}
\caption{(a) 2-12 keV ASM lightcurve and (b) hardness ratio (5-12 keV vs. 2-5 keV count ratio) 
as a function of the MJD of the observation. The vertical dashed lines at MJD 55296, MJD 55304, and 
MJD 55331 indicate the state transitions from hard to hard-intermediate state, then to soft-intermediate 
state and finally to soft state.}}
\label{kn : fig1}
\end{figure}

\begin{figure}
\vbox{
\vskip 0.3cm
\centerline{
\includegraphics[scale=0.6,angle=0,width=7truecm,height=5truecm]{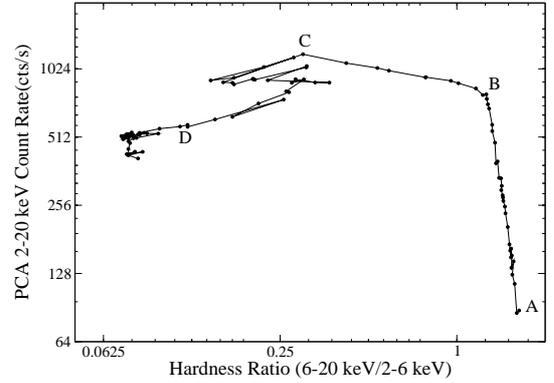}}
\vspace{0.0cm}
\caption{Hardness intensity diagram (HID) of GX 339-4 observed with RXTE/PCA from 2010 January 12 
to August 14, as it approaches the soft state from the hard state via hard-intermediate and 
soft-intermediate spectral states. The total count rates in the 2-20 keV energy band along 
Y-axis and the ratio of the count rates in the 6-20 keV and 2-6 keV bands are given on the X-axis. 
The points A, B, C, and D are on MJD 55208, MJD 55296, MJD 55304, and MJD 55331 respectively. 
Point A indicates our first observation day and points B, C, and D indicate the state transitions.}}
\label{kn : fig2}
\end{figure}

\begin {figure}[t]
\vskip 0.8 cm
\centering{
\includegraphics[scale=0.6,angle=0,width=8truecm]{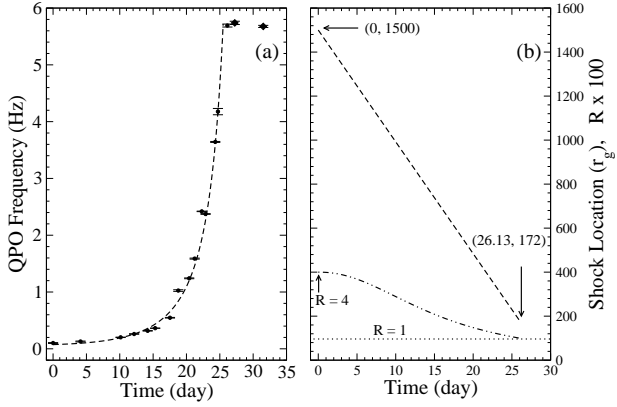}}
\caption{(a) Variations of the QPO frequency with time (in day) of the rising phase of the 
outburst with the fitted POS model (dotted curve) are shown. The diamonds indicate the last 
two incidents of observed QPOs (on 2010 April 18 and 22) and they are not included in our 
model fitting, because the shock already achieved its weakest strength of $R\sim 1$ earlier 
on April 17. (b) Variation of the shock location (in $r_g$) and shock strength ($R$).
The shock seems to be stagnating at around $172$ Schwarzschild radii.}

\label{kn : fig3}
\end {figure}

\begin{figure}
\vbox{
\vskip 0.5cm
\centerline{
\includegraphics[scale=0.6,angle=0,width=8truecm]{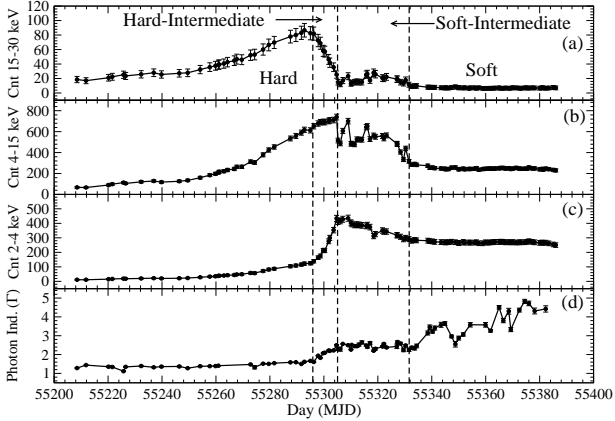}}
\vspace{0.0cm}
\caption{Variation of the RXTE/PCA count rates in (a) 15-30 keV, (b) 4-15 keV and (c) 2-4 keV 
energy bands and (d) the power-law photon index $\Gamma$ with day (in MJD) are plotted. The 
vertical dashed lines are on MJD 55296, MJD 55304.7 and MJD 55331, indicating the state 
transitions from hard to hard-intermediate, to soft-intermediate and to soft states respectively.}}
\label{kn : fig4}
\end{figure}


\section{Discussions and concluding remarks}

We analyzed the recent outburst of the black hole candidate GX 339-4. We studied the evolution 
of the spectral and timing properties since 2010 January 12 till 2010 July 8 and showed that
there was no signature of QPO till March 21. After this, until April 17, QPO was continuously 
observed. The frequency was monotonically increasing in a very similar way to what was shown 
in GRO J1655-40 and XTE J1550-564. Until April 9, the spectral photon index was less than $1.6$, 
the source was at a pure hard state. Then the object moved to the hard-intermediate state as 
the Keplerian rate started increasing when the photon spectral index changed from $\sim 1.6$ 
to $\sim 2.0$. This short duration state was continued until April 17. Then as the Keplerian 
rate became comparable to the sub-Keplerian rate, the spectral index rapidly became soft 
($\Gamma \sim 2.5$) and the object remained in the soft-intermediate state for $\sim 26$ days 
until 2010 May 14. Here sporadic QPOs were observed and the QPO frequency remained at around 
$6$ Hz. After that the Keplerian rate dominates and the spectrum becomes softer with a high 
spectral index ($\Gamma > 3.0$). We believe that when the viscosity is reduced in future, the 
Keplerian rate will be so much reduced that the Keplerian disk may itself evaporate and the 
inner edge of the Keplerian could be thought to have receded from the black hole. In this 
case, the object would return to the hard state (Mandal \& Chakrabarti, 2010) and monotonically 
decreasing nature of the QPO frequencies will be observed  as predicted by the POS solution 
(CDNP08, CDP09).

In Chakrabarti \& Manickam (2000), the authors showed that the true `soft photons' i.e., the
photons emitted from the pre-shock flow, do not exhibit QPOs. A cross-check that the two-component 
model is valid and that the QPO is caused by the oscillation of the Comptonized cloud (namely, 
the post-shock region) can be made. As the shock moved in, the post-shock flow was heated up 
roughly as $T\sim 1/r$ while the pre-shock Keplerian flow was heated up as $ r^{-3/4}$ 
(Shakura \&  Sunyaev, 1973). This means that so-called `soft-photons' in $2-4$ keV range 
could be actually Comptonized photons at a large distance and should be treated 
as hard photons. Thus, $2-4$ keV photons should show QPOs at a large distance only.
In Fig. 5(a-c) we plot the power density spectra of photons in the $2-4$ keV, $4-15$ keV and
$15-30$ keV bands on the April 5, 13 and 16 respectively when the shock (Compton cloud boundary) 
was located at $\sim 780 r_g, 367 r_g$ and $244 r_g$ respectively with compression ratios of 
$2.2$, $1.27$ and $1.08$ respectively. The spectral and timing properties of the object in 
and around these days are presented in Table 1 and Table 2. In Table 1 we show how the photon 
index and fluxes vary in these days. $\chi^2$ and the number of degrees of freedom (dof)
in the spectral fits are also included. In Table 2 we note that the rms amplitude of the 
fundamental QPO decreased with time in $2-4$ keV range. This shows that the number of 
Comptonized photon is decreasing in this energy bin and increasing in the $4-15$ keV 
bin as the shock moves in. We note that in the $4-15$ keV range, the rms initially decreased 
and then increased. The latter increase is because the entire Comptonized photons now 
belong to this energy bin. In high-energy channels ($15-30$ keV) the rms amplitude 
decreased because of the paucity of photons. 

Clearly Table 2 shows that the rms amplitude was highest in all energy ranges when 
the shock was farther out. This shows that the oscillation of the Compton cloud, which 
was caused by a stronger shock at a large distance, was high enough to significantly 
modulate the outgoing flux. This is particularly surprising in the $2-4$ keV range, 
because it was supposed to be the energy of the `soft-photons'. This shows that far out, 
these photons are actually Comptonized photons. Note that the QPO is absent in this 
range on April 16. This is expected, because the Compton cloud is so close to the 
black hole that these photons are contributed mostly from the Keplerian flow and are 
not modulated. Note that the appearance of harmonics (Yu, 2010) depends on the energy 
of the photons. On the April 16, the shock is very weak and a faint QPO is seen only 
in $4-15$ keV. At lower energy bins QPO was not seen because they were emitted from the 
pre-shock flow, and at higher energy bins, the number of photons was statistically 
insignificant. This analysis indicates that the general two-component solution of 
the outburst sources as developed in CDNP08, CDP09, and Mandal \& Chakrabarti (2010) 
can explain most of what is seen in GX 339-4 so far. Assuming that this outburst is 
similar to that in GRO J1655-40 (CDNP08) we expect that once the object goes to the 
low-hard state, it will take about 32-38 days to return to quiescence and the QPO
frequency would monotonically decrease from $\sim 6$Hz to a few mHz in that period.  

Miller et al. (2006), Ramadevi \& Seetha (2007), and Rykoff et al. (2007) recently 
argued that the presence of a thermal emission even in the low-hard state of some 
outburst sources points to a Keplerian disk with an inner edge very 
close to the black hole. The objects exhibit a behavior similar to a classical 
outburst with FRED-type lightcurve. In these cases, the data in the rising phase 
are scarce and these authors used the data from the day when the outburst was near 
its peak and had a softer spectrum. The conclusions of these authors do not 
necessarily mean that the Keplerian disk cannot move in within the viscous time 
scale as in our paradigm. According to our picture, the Keplerian disk has already 
moved in during the rising part and was present when these authors commenced their 
analysis.

\begin{table}[h]
\scriptsize
\centering
\caption{\label{table1} Spectral properties during the initial outburst phase} 
\vskip 0.0cm
\begin{tabular}{|l|c|c|c|c|c|}
\hline
Obs. Id$^*$ & UT &Photon & \multicolumn{2}{|c|}{Flux$^\dagger$}&$\chi^2$/dof$^{**}$  \\
\cline{4-5}
       & Date    &Index($\Gamma$)&3-10 keV & 10-25 keV &   \\
\hline
X-10-05&2010-03-17&1.473&2.533&2.868&39.92/48 \\
\hline
X-13-02&2010-04-05&1.504&2.405&2.545&55.13/48 \\
\hline
X-14-06&2010-04-13&2.073&6.495&3.637&43.84/48 \\
\hline
X-15-00&2010-04-16&2.185&3.446&1.390&68.51/48 \\
\hline
X-16-04&2010-04-28&2.613&6.238&0.658&56.74/48 \\
\hline
\end{tabular}
\leftline {$^\dagger$ Flux in unit of $10^{-9} ergs~cm^{-2}~s^{-1}$}
\leftline {$^*$ Here, X=95409-01. $^{**}$ dof means no. of degrees of freedom}
\end{table}

\begin{table}[h]
\scriptsize
\centering
\caption{\label{table1} Timing properties during the initial outburst phase} 
\vskip 0.0cm
\begin{tabular}{|l|c|c|c|c|}
\hline
Obs. Id$^*$ & UT & \multicolumn{3}{|c|}{$\nu_{QPO}$ (Hz) and rms amp.(\%)} \\
\cline{3-5}
       & Date    &2-4 keV & 4-15 keV & 15-30 keV  \\
\hline
X-10-05&2010-03-17& --- & --- & --- \\
\hline
X-13-02&2010-04-05& 0.309,  15.190   & 0.316, 16.048  & 0.313, 11.893  \\
\hline
X-14-06&2010-04-13& 2.420, \  7.095  &2.424, \ 8.258 & 2.430, \ 7.292 \\
       &     & 4.846, \  6.056     &  4.813, \ 5.585    & ---\\
       &     & 7.139, \  4.241     &  7.310,\ 4.987    &  --- \\
\hline
X-15-00&2010-04-16& --- & 4.153, \ 10.293  & ---\\
\hline
X-16-04&2010-04-28& ---  & --- &--- \\
\hline
\end{tabular}
\leftline {$^*$Here, X=95409-01.}
\end{table}

Black hole accretion is a complex process, and it is abundantly clear that a 
simple standard disk (Shakura \& Sunyaev, 1973) is not capable of explaining 
most of the observations. Out of all the observations, the outbursting sources 
are extremely important, because they exhibit the changes in spectral states 
in rapid succession. Similarly they also generally exhibit a systematic variation 
of QPO frequencies. These enable us to study the dynamics of matter close to a 
black hole. A number of outbursting sources, whether they exhibit FRED lightcurves 
as in the soft X-ray transients or slow-rise and slow-decay (SRSD) lightcurves 
as in GX 339-4 or GRO J1655-40 show similar and timing properties. Our 
two-component flow paradigm seems to be capable of explaining these sources quite 
naturally if one assumed that the outburst is caused by a rapid rise of viscosity, 
which drives Keplerian flows towards the black hole and/or converts some of the 
low angular momentum flows into Keplerian flows. Because quantifying viscosity 
is not easy, a prediction of a detailed behavior has not been possible so far. 
However, analyses of these sources are very useful in advance our quest for a 
general solution of this difficult problem. 

\begin{figure}
\vbox{
\vskip -0.0cm
\centerline{
\includegraphics[scale=0.6,angle=0,width=4.5truecm]{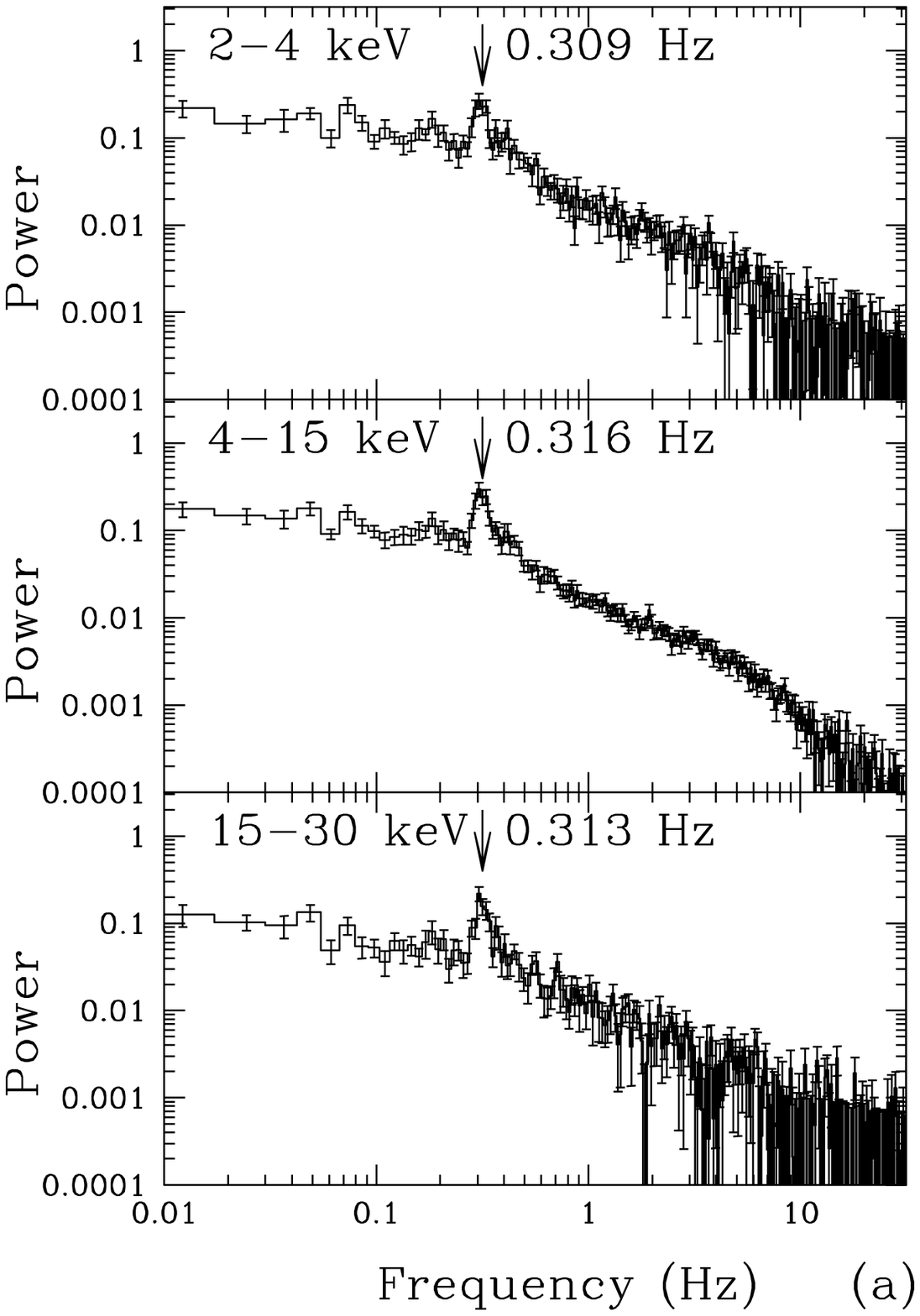}\hskip -1.5cm
\includegraphics[scale=0.6,angle=0,width=4.5truecm]{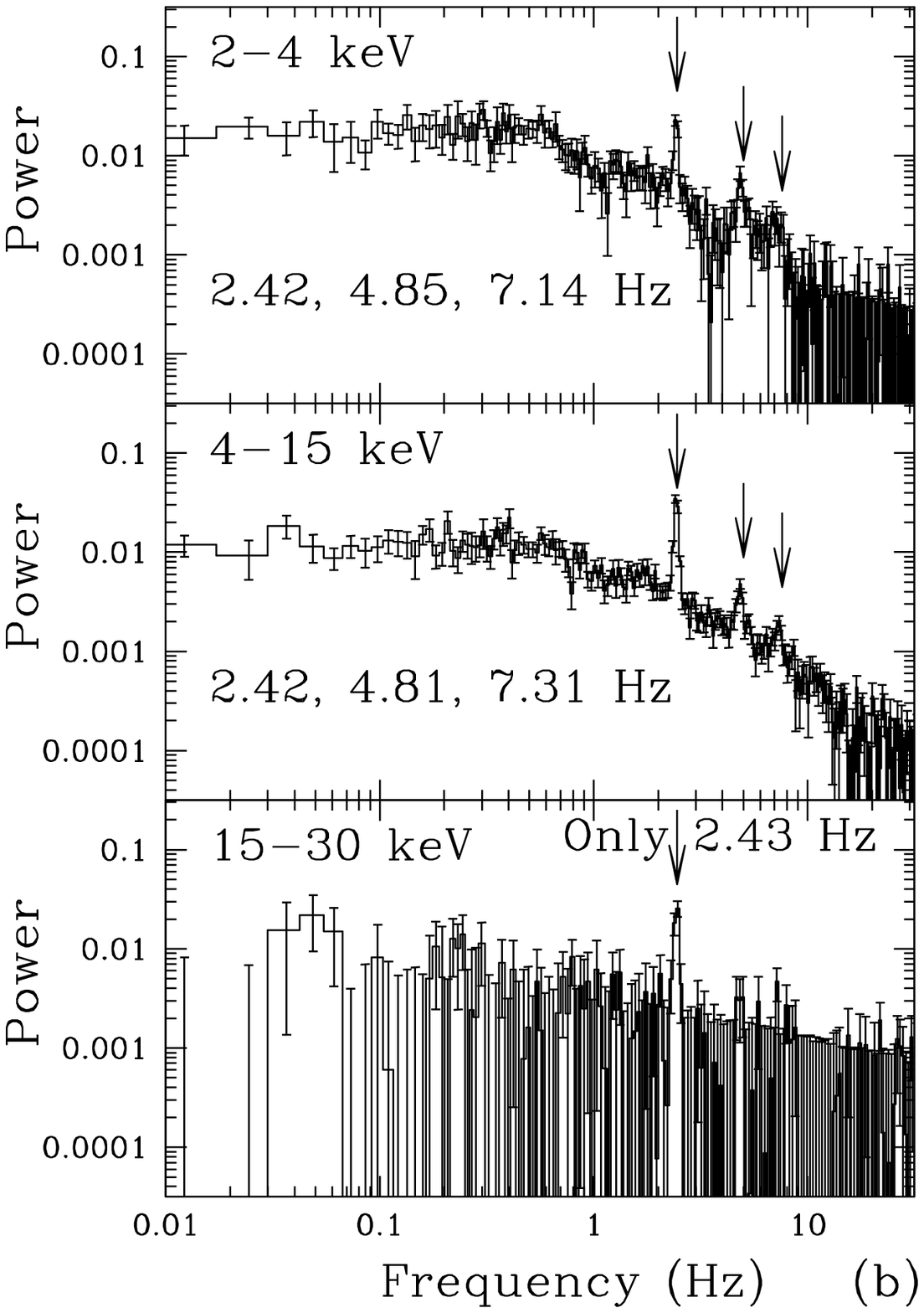}\hskip -1.5cm
\includegraphics[scale=0.6,angle=0,width=4.5truecm]{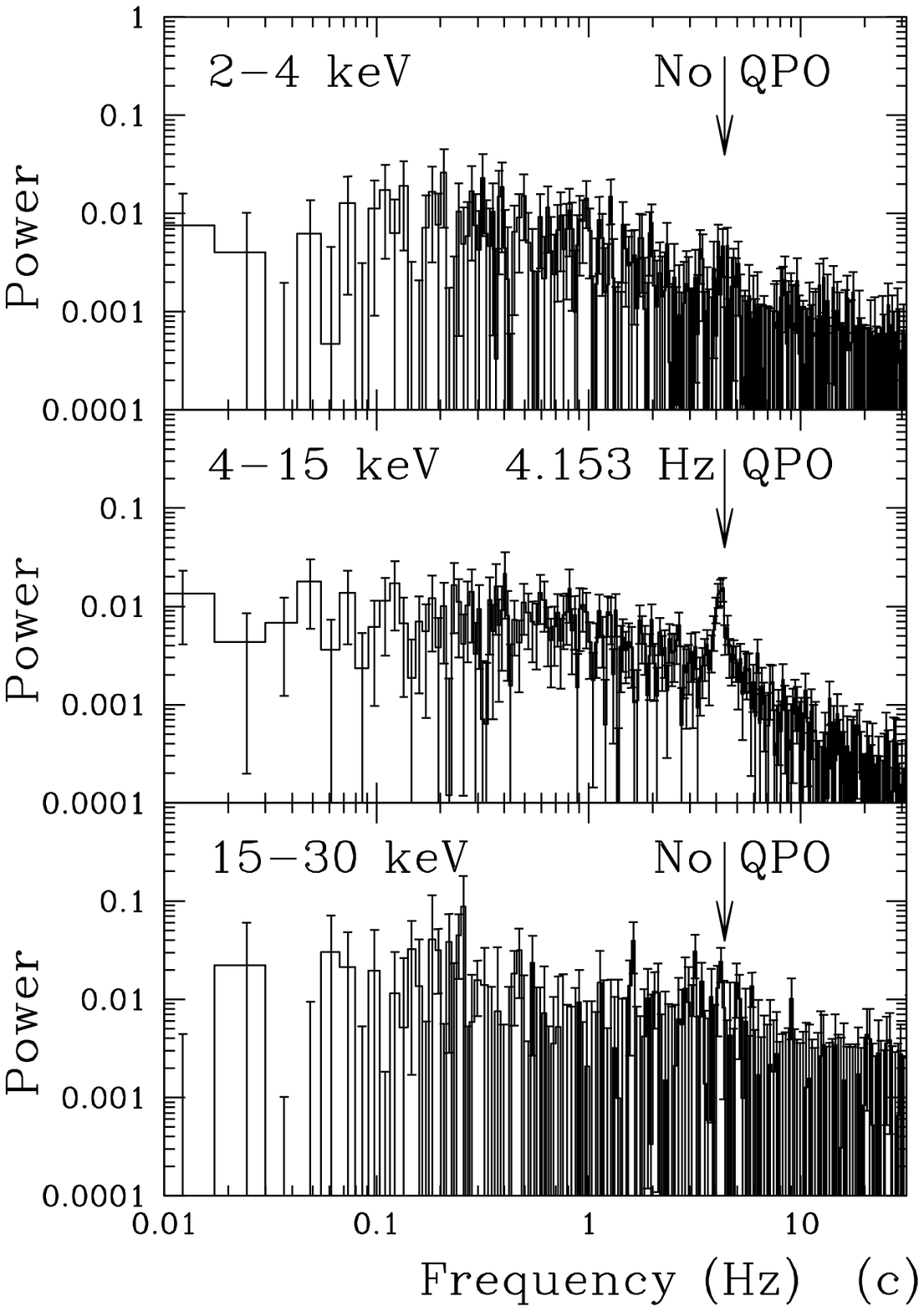}
}
\vspace{0.0cm}
\caption{(a-c): Power density spectrum of photons in three different energy bands in three 
different observational IDs: (a) Obs. ID:95409-01-13-02 of 2010 April 5, (b) Obs. 
ID:95409-01-14-06 of 2010 April 13, and (c) Obs ID:95409-01-15-00 of 2010 April 16. The top, 
middle, and bottom panels are for $2-4$ keV, $4-15$ keV, and $15-30$ keV bands respectively. 
Shock locations are found to be at $780 r_g$, $367 r_g$, and $244 r_g$, and the compression 
ratios (which represent the shock strength) are $2.20$, $1.27$, and $1.08$ respectively. 
In (a) $2-4$ keV photons are actually Comptonized photons, they also show strong QPOs, while 
in (c) they are soft photons and do not show QPOs (Chakrabarti \& Manickam, 2000). Note also 
that because the shock is stronger in (a) the {\it rms} amplitude is higher in all three 
energy bands, while in (c) the shock is weaker and the {\it rms} is also very low. In (b), the
fundamental and two harmonics could be seen.}}
\label{kn : fig5}
\end{figure}

\section*{Acknowledgments}

D. Debnath acknowledges the support of CSIR-NET scholarship.

{}

\end{document}